\documentclass[aps,prl,reprint,groupedaddress,amsmath,amssymb,longbibliography,floatfix]{revtex4-2}

\usepackage{graphicx}
\usepackage{placeins}
\usepackage{dcolumn}
\usepackage{bm}
\usepackage{xcolor}
\usepackage{epsf}
\usepackage{epsfig}

\usepackage{hyperref}

\begin{document}
\title{Chiral Electron Momentum Distribution upon Strong-Field Ionization of Atoms}

\author{A. Geyer$^1$}
\email{geyer@atom.uni-frankfurt.de}
\author{J. Stindl$^1$}
\author{I. Dwojak$^1$}
\author{M. Hofmann$^1$}
\author{N. Anders$^1$}
\author{P. Roth$^1$}
\author{P. Daum$^1$}
\author{J. Kruse$^1$}
\author{S. Jacob$^1$}
\author{S. Gurevich$^1$}
\author{N. Wong$^1$} 
\author{M. S. Schöffler$^1$}
\author{L. Ph. H. Schmidt$^1$}
\author{T. Jahnke$^2$}
\author{M. Kunitski$^1$}
\author{R. Dörner$^1$}
\author{S. Eckart$^1$}
\email{eckart@atom.uni-frankfurt.de}

\affiliation{$^1$ Institut f\"ur Kernphysik, Goethe-Universit\"at, Max-von-Laue-Str. 1, 60438 Frankfurt am Main, Germany}
\affiliation{$^2$ Max-Planck-Institut f\"ur Kernphysik, Saupfercheckweg 5, 69117 Heidelberg, Germany}

\date{\today}
\begin{abstract}
We present a scheme to synthesize a three-dimensional laser field that produces a chiral electron momentum distribution upon strong-field ionization of atoms. Our approach employs two orthogonally propagating two-color laser beams. This results in a time-dependent three-dimensional electric field vector of the combined light field which varies for different positions within the focal volume. For each position, we conduct a simulation of the corresponding electron momentum distribution that includes non-adiabatic dynamics and Coulomb interaction after tunneling. For suitable laser parameters, only a small region of the focal volume contributes to the final momentum distribution. Thus, integrating over all position coordinates, a specific chiral laser field dominates. This leads to a volume-averaged electron momentum distribution, which is chiral, as well. This work will serve as a benchmark for future strong-field experiments aiming at the synthetization of well-defined, three-dimensional laser fields.
\end{abstract}

\maketitle

\section{I. Introduction}

Strong-field ionization can occur when an atom or molecule is irradiated by a strong laser field \cite{Keldysh1965}. This process strongly depends on the symmetries of the driving light field. Pioneering work on strong-field ionization was performed using linearly, circularly, and elliptically polarized light \cite{agostini1979p, bucksbaum1986suppression, Corkum1989, krauseetal_cutofflaw, Eckle2008, Hanus2019, Meckel2008, weber2000correlated,Staudte2009,Arissian2010}. In 1999, Becker et al. suggested using two-color light fields to generate elliptically polarized high harmonics \cite{Becker1999}. Later these fields were also successfully used to study ionization processes \cite{MancusoPRA2015_CRTC, Mancuso2017_enhancement, fleischer2014spin}. This variety of experimentally available laser electric fields enabled the study of a wide range of phenomena, such as the production of spin-polarized electrons, m-selective tunneling, and Wigner time delays \cite{Barth2011,Herath2012,Barth2013,Eckart2018_Offsets,EckartNatPhys2018,trabert2021angular,Han2018}. However, all experiments on strong-field ionization had in common that they were driven by laser fields with a time-dependent electric field that was restricted to either a line (e.g. linearly polarized light) or a plane (e.g. circularly polarized light). We refer to these fields as one-dimensional (1D) and two-dimensional (2D) light fields, respectively. 

The reduced dimensionality causes a major limitation for studies of strong-field ionization in the tunneling regime. This is because tunneling acts like a filter and liberates a part of the bound electronic wave function close to the tunnel exit position \cite{Arissian2010,Murray2010,trabert2021angular}. Since the tunnel exit position is governed by the direction of the electric field vector, all possible tunnel exit positions are restricted to be close to a line for 1D light fields or a plane for 2D light fields. This restriction makes it difficult to access three-dimensional properties of atoms and molecules when using 1D or 2D light fields. 

The goal of the present paper is to overcome this limitation of dimensionality and suggest a scheme allowing the study of ionization by three-dimensional (3D) light fields. The main challenge to be met is that the desired three-dimensionality of the time-dependent field vector comes at the price of a strong position dependence of the 3D laser electric field across the focal volume. 

Pioneering theoretical work proposing 3D light fields include using non-colinear laser beams \cite{neufeld2021strong, ayuso2019synthetic, Ofer2019_Ultrasensitive_Chiral_Spectroscopy, Ofer2020,Mayer2022}, or vortex light \cite{mayer2024chiral} to synthesize 3D light fields. Those 3D light fields could e.g. be used for the detection of enantiosensitive observables \cite{mayer2024chiral}, all-optical enantiopurification \cite{neufeld2021strong}, the generation of chiral atoms \cite{Mayer2022}, ultrasensitive chiral spectroscopy based on high-harmonic generation in 3D light fields \cite{Ofer2019_Ultrasensitive_Chiral_Spectroscopy}, or to study light-matter interaction in three dimensions in general \cite{habibovic2024emerging}. Previous theoretically proposed methods did not allow for the synthesization of light fields that lead to chiral electron momentum distributions upon strong-field ionization of an achiral target (e.g. a rare gas atom) \cite{ayuso2019synthetic, neufeld2021strong, Rego_2023}. Although former studies have shown that it is possible to have one dominating handedness of the electric field's Lissajous curve over the entire focal volume (see ``globally chiral light'' \cite{ayuso2019synthetic}), the Lissajous curve of the laser electric field and the negative vector potential were position dependent in previous work and thus they varied across the focal volume.  Since the negative vector potential significantly determines the electron's dynamics after tunneling \cite{Corkum1993,Eckle2008,Shilovski2016,Ni2018_theo}, such a light field leads to an electron momentum distribution, which has properties that are not related to a single negative vector potential but to a distribution of vector potentials as in experiments one inevitably integrates over the entire focal volume.

\begin{figure}[t!]
\includegraphics[width=\columnwidth]{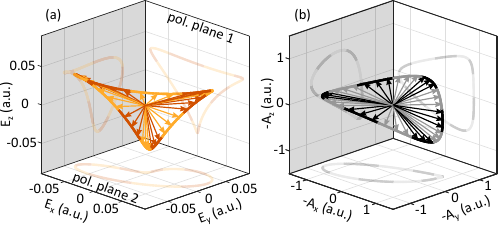}
\caption{\label{fig1} Three-dimensional laser electric field and corresponding negative vector potential. (a) shows the Lissajous curve of a three-dimensional (3D) laser electric field $\vec{E}(t)$ in light and dark orange and its projections. The arrows show the electric field vector at different instants of time. The two alternating colors mark the twelve time intervals into which the field is divided for the simulation. (b) shows the corresponding negative vector potential $-\vec{A}(t)$ in black and gray in full analogy to (a).}
\end{figure}

In contrast, in this paper, we present a scheme for creating a well-defined globally chiral laser field that produces a chiral electron momentum distribution upon strong-field ionization of an argon atom in the ground state. Here, ``well-defined'' refers to a situation where only one electric field (with one defined handedness) and the corresponding negative vector potential dominate the ionization process. The measurable electron momentum distribution will serve as a test case to evaluate whether the driving light field is indeed a well-defined 3D light field. Such a driving field is ideally suited to study a new domain of chiral light-matter interaction. While previously the chirality was determined by the target, for this new class of laser fields the chirality of the driving light field governs the interaction. In our proposed approach the combined electric field and the combined negative vector potential are both chiral and have a well-defined handedness even within the dipole approximation. 

To prove the feasibility of the novel scheme, we conduct a numerical simulation which combines the strengths of strong-field approximation (SFA) and a classical two-step (CTS) model, which is well established for the simulation of strong field ionization in 2D light fields \cite{Brennecke2020_gouy,Trabert2021Atomic,Geyer2023Experimental,hofmann2024subcycle}. The simulation method is described in section II. In section III the properties of the globally chiral laser field that is used to generate chiral electron momentum distributions are described. The globally chiral laser field is generated by overlapping two perpendicularly propagating two-color laser beams. Due to the relative phase of the two laser pulses, many different 3D Lissajous curves for the laser's electric field exist within the focal volume, all contributing to the measured electron momentum distribution. In section IV we will show that, due to the highly nonlinear nature of tunneling as a function of the magnitude of the laser field, it is possible to choose field parameters that ensure that only those regions in the focal volume contribute to the ionization signal where the 3D light field has a well-defined handedness. This leads to a chiral electron momentum distribution based on realistic experimental conditions. Eventually, the chirality of these momentum distributions will be quantified using a scalar-valued figure of merit.\\

\section{II. Simulation method}\label{sec:simulation}
\label{secsimulation}
\begin{figure}[t!]
\includegraphics[width=\columnwidth]{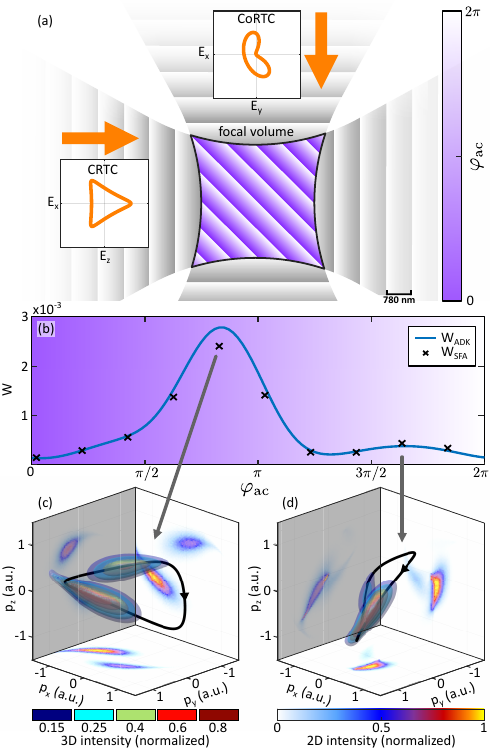}
\caption{\label{fig3} Position-dependent three-dimensional laser fields from perpendicularly propagating CRTC and CoRTC fields. (a) shows a schematic sketch of the two laser beams crossing. One of the laser beams is a counter-rotating two-color (CRTC) field and the second beam is a co-rotating two-color (CoRTC) field as indicated by the insets in orange. In the overlapping region (focal volume) 3D laser fields emerge. The relative phase of the CRTC and CoRTC field $\varphi_{ac}$ is position-dependent (purple colorbar). The combined 3D electric field and the resulting ionization probability depend on $\varphi_{ac}$. (b) shows the approximated ionization probability $W$ for one optical cycle as a function of $\varphi_{ac}$ calculated with the ADK ionization rate (blue line) and the rate based on SFA (black crosses). (c) [(d)] shows the negative vector potential in black for $\varphi_{ac}=2.6$ [$\varphi_{ac}=5.1$] and the corresponding 3D final electron momentum distribution calculated with the combined SFA and CTS model. (c) shows the same negative vector potential as in Fig. \ref{fig1}(b). Please note that (c) and (d) both have two shared colorbars, one for the 3D isosurfaces and one for the 2D projections. Due to normalization, the maximum value of 1 is only reached in a very small volume (in only one or a few bins of the histogram). To ensure that the isosurfaces represent a larger (visible) volume, the highest value of the discrete scale is chosen to be 0.8.}
\end{figure}

Fig. \ref{fig1} shows a 3D laser electric field $\vec{E}(t)$ and the corresponding negative vector potential $-\vec{A}(t)$. Here, $\vec{E}(t)$ is the superposition of a counter-rotating two-color (CRTC) laser field propagating in $y$-direction (polarization plane 1: $E_xE_z$-plane) and a co-rotating two-color (CoRTC) laser field propagating in $z$-direction (polarization plane 2: $E_xE_y$-plane). The Lissajous curves of the CRTC and the CoRTC light field are shown in Fig. \ref{fig3}(a). The 3D laser field shown in Fig. \ref{fig1} is used throughout this paper. $\vec{E}(t)$ is given by Eq. \ref{field}. Here $E_{tc1,\omega}$, $E_{tc1,2\omega}$, $E_{tc2,\omega}$ and $E_{tc2,2\omega}$ are the field maxima of the single colors, $e_{tc1,\omega}$, $e_{tc1,2\omega}$, $e_{tc2,\omega}$ and $e_{tc2,2\omega}$ are the ellipticities of the single colors, $\varphi_{tc1,\omega}$, $\varphi_{tc1,2\omega}$, $\varphi_{tc2,\omega}$ and $\varphi_{tc2,2\omega}$ are the phase offsets of the single colors, $\varphi_{tc1}$ and $\varphi_{tc2}$ are the phase offsets of the $2\omega$ fields relative to the $\omega$ fields and $\varphi_{ac}$ is the phase offset between the two beams that propagate in orthogonal directions \footnote{The values for the electric field shown in Fig. \ref{field}(a) are: $E_{tc1,\omega}=0.045$ a.u., $E_{tc1,2\omega}=0.015$ a.u., $E_{tc2,\omega}=0.043$ a.u., $E_{tc2,2\omega}=0.015$ a.u., $e_{tc1,\omega}=1$, $e_{tc1,2\omega}=1$, $e_{tc2,\omega}=0.4402$, $e_{tc2,2\omega}=1$, $\varphi_{tc1,\omega}=0$, $\varphi_{tc1,2\omega}=\pi$, $\varphi_{tc2,\omega}=5.5551$, $\varphi_{tc2,2\omega}=0$, $\varphi_{tc1}=4.7124$, $\varphi_{tc2}=1.5136$, $\varphi_{ac}=2.6044$.}.

In order to simulate the electron momentum distribution that emerges upon strong-field ionization of an atom using the 3D light field from Fig. \ref{fig1}, we use the following approach: In the first step an electron is liberated through tunnel ionization. The tunneling probability, the initial electron momentum distribution and the tunnel exit position are determined by SFA in the length gauge \cite{Popruzhenko2008, Yan_2012}. In the second step the electron propagates classically in the combined potential of the remaining ion and the laser field (CTS model) using the initial conditions from SFA \cite{Shilovski2016}. Since SFA is usually defined within a single polarization plane, we divide $\vec{E}(t)$ into small time intervals for the 3D case (here twelve time intervals are used, see Fig. \ref{fig1}). To test the convergence of this approach, we varied the number of time intervals using up to 100 time intervals and did not find relevant deviations. For each time interval the electric field can be approximated by a 2D light field lying in a plane. These planes are referred to as instantaneous polarization planes and are defined by two vectors: The electric field vector in the middle of the corresponding time interval and its time derivative. For each time interval the electric field is approximated as the superposition of two elliptically polarized fields with frequencies $\omega$ and $2\omega$.\parfillskip=0pt
\onecolumngrid
\begin{align}\label{field}
E_x(t)&= E_{tc1,\omega}\cdot \cos(\omega t)+E_{tc2,\omega}\cdot e_{tc2,\omega}\cdot \sin(\omega t+\varphi_{ac}+\varphi_{tc2,\omega})+\notag \\
&E_{tc1,2\omega}\cdot \cos(2\omega t+\varphi_{tc1})+E_{tc2,2\omega}\cdot e_{tc2,2\omega}\cdot \sin(2\omega t+2\varphi_{ac}+\varphi_{tc2}+\varphi_{tc2,2\omega})\notag \\
E_y(t)&= E_{tc2,\omega}\cdot \cos(\omega t+\varphi_{ac})+E_{tc2,2\omega}\cdot \cos(2\omega t+2\varphi_{ac}+\varphi_{tc2})\notag \\
E_z(t)&= E_{tc1,\omega}\cdot e_{tc1,\omega}\cdot \sin(\omega t+\varphi_{tc1,\omega})+E_{tc1,2\omega}\cdot e_{tc1,2\omega}\cdot \sin(2\omega t+\varphi_{tc1}+\varphi_{tc1,2\omega})
\end{align}
\twocolumngrid
\noindent  The intensities, ellipticities, orientation of the elliptical axis,  helicities and the relative phase of the $\omega$ and $2\omega$ polarization ellipses are fitted to approximate the fields in the respective segment of the 3D electric field's Lissajous figure. This approximated 2D field is then used to obtain the initial conditions of the electrons at the tunnel exit from SFA. The electrons are then propagated classically in the presence of the time-dependent 3D laser field and the Coulombic potential. Integrating the results of this procedure for all time intervals within one cycle of the driving light field results in the electron momentum distribution shown in Fig. \ref{fig3}(c). We have calculated one million trajectories for this distribution. To test for convergence, we redid the simulation for Fig. \ref{fig3}(c) using 50 million trajectories and did not find relevant differences. For the sake of completeness, we have analyzed the percentage of electrons that recollide with their parent ion to be $0.5\%$ (determined by counting the number of trajectories that approach their parent ion to within 5 a.u.) for the electric field discussed in this work.

\section{III. Position-dependent 3D laser field} \label{sec:position}

\begin{figure*}[ht]
\includegraphics[width=\textwidth]{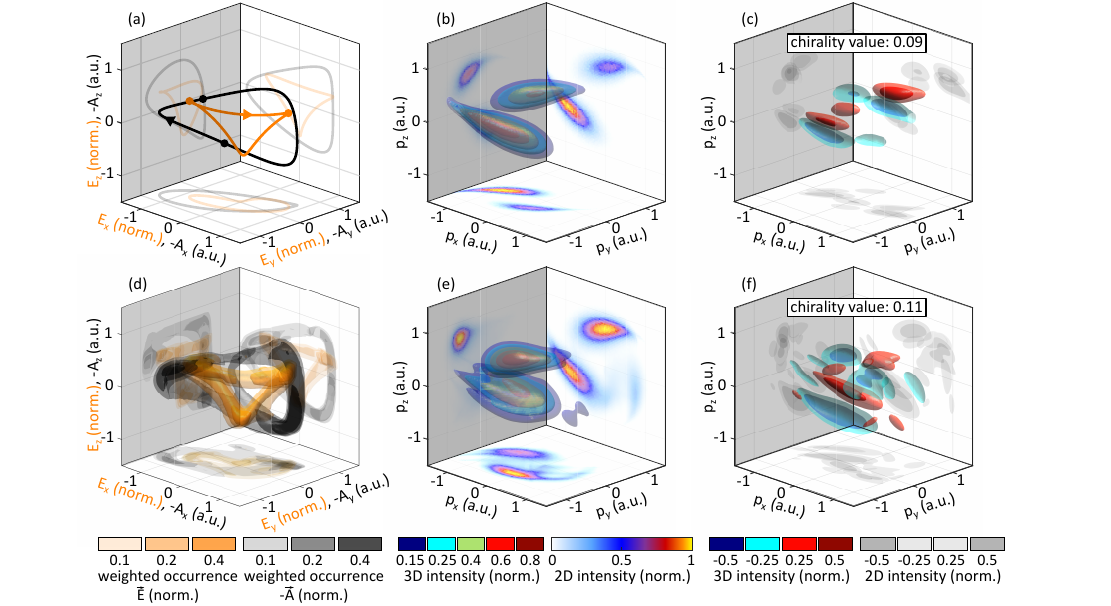}
\caption{\label{fig4} Chiral electron momentum distributions. (a) shows the same laser electric field and negative vector potential as Fig. \ref{fig1}(a) and \ref{fig1}(b). The orange dots mark the two peak electric fields and the black dots the corresponding negative vector potentials. The arrows indicate the temporal evolution. (b) shows the final electron momentum distribution for the laser field shown in (a), which is the 3D field with the highest ionization probability in the studied scenario (see $\varphi_{ac}=2.6$ in Fig. \ref{fig3}(b)). (c) illustrates that the distribution in (b) is chiral (see text for details and note that for an achiral distribution the histogram would be empty and the chirality value would be zero). (d) illustrates that there is a range of 3D fields contributing to ionization in full analogy to (a) if the entire focal volume contributes. To this end $\vec{E}(t)$ is calculated for all values $\varphi_{ac}$ that are indicated in Fig. \ref{fig3}(b). The electric field and the negative vector potential are filled into the histogram in (d) and weighted with the maximum ADK rate of the combined electric field. (e) shows the total final electron momentum distribution, which is the sum of all electron momentum distributions for the ten values of $\varphi_{ac}$. (f) shows the same as (c) for the distribution in (e). It is evident that (e) is a chiral electron momentum distribution.}
\end{figure*}


To produce 3D laser fields in the lab one can use non-colinearly propagating laser pulses. Fig. \ref{fig3}(a) shows a schematic sketch of a CRTC laser beam and a CoRTC laser beam crossing. In the overlapping region (focal volume) a 3D laser field emerges. The purple colorbar indicates the relative phase of the two beams $\varphi_{ac}$ (see Eq. \ref{field}), which is position-dependent. In Fig. \ref{fig1} the value of $\varphi_{ac}$ is set to $2.6$. However, for different values of $\varphi_{ac}$ the shape of the laser field and negative vector potential vary. We will overcome this limitation by exploiting that in our scheme the approximated ionization probability $W$, integrated over one optical cycle, varies as a function of $\varphi_{ac}$, which can be used for an improved intensity gating \cite{ayuso2019synthetic}.  The electron momentum is mainly determined by the negative vector potential. Thus, the key to generating chiral electron momentum distributions is to find a scenario where the ionization rate is dominated by regions of a certain relative phase for which the negative vector potential shows a chiral Lissajous figure. The scenario shown in Fig. \ref{fig1} is one example which satisfies these conditions as we demonstrate in Fig. \ref{fig3}(b). Here $W$ is shown as a function of $\varphi_{ac}$. $W$ is evaluated using both the ADK $\omega_{\mathrm{ADK}}$ \cite{Ammosov1986, Shilovski2016} and SFA $\omega_{\mathrm{SFA}}$ approaches and is calculated by integrating the rate over one full cycle of the 3D laser field:
\begin{equation}
W(\varphi_{ac})=\int_0^T \omega_{\mathrm{ADK/SFA}}(\vec{E}(t,\varphi_{ac}))dt\ .
\end{equation}
It is found that $\varphi_{ac}=2.6$ has the highest contribution to the overall ionization probability per optical cycle. Fig. \ref{fig3}(c) shows the negative vector potential in black for $\varphi_{ac}=2.6$ and the corresponding 3D final electron momentum distribution calculated with the combined SFA and CTS model. Fig. \ref{fig3}(d) shows the same as Fig. \ref{fig3}(c) but for $\varphi_{ac}=5.1$. It is evident that choosing a different $\varphi_{ac}$ can modify the negative vector potential and the final electron momentum distribution significantly. Even though both momentum distributions from Fig. \ref{fig3}(c) and \ref{fig3}(d) contribute to the total momentum distribution, it is evident that the distribution shown in Fig. \ref{fig3}(c) clearly dominates due to a higher ionization probability.

\section{IV. Chiral electron momentum distribution}\label{sec:ChiralPEMD}

Fig. \ref{fig4}(a) shows the same laser electric field and negative vector potential as in Fig. \ref{fig1}(a) and \ref{fig1}(b). Fig. \ref{fig4}(b) shows the corresponding final electron momentum distribution (the same as in Fig. \ref{fig3}(c)). To prove that the electron momentum distribution displayed in Fig. \ref{fig4}(b) is chiral, we introduce a figure of merit that quantifies chirality. This measure of chirality is closely related to the method in \cite{zabrodsky1995continuous}. To this end the original distribution from Fig. \ref{fig4}(b) is referred to as $P$. Then, $P$ is normalized to fulfill $\int P\ dp_xdp_ydp_z=1$ and the center of gravity of the distribution is taken as the new origin. $\tilde{P}$ is the point-mirrored version of $P$. Next, $\tilde{P}$ is rotated around the x-, y- and z-axis (rotation angles $\varphi$, $\theta$ and $\eta$) and the resulting distribution is referred to as $\tilde{P}_\mathrm{rot}(\varphi,\theta,\eta)$. If $P$ were chiral, it cannot be superimposed on its mirror image  $\tilde{P}_\mathrm{rot}(\varphi,\theta,\eta)$, even after rotation or translation. In contrast, if $P$ were achiral, then $\tilde{P}$ can be rotated in a way, that $P$ and $\tilde{P}_\mathrm{rot}(\varphi,\theta,\eta)$ are identical. The chirality value $\mu_\textrm{CV}$ of $P$ is calculated via:
\begin{align}
\mu_\textrm{CV} = \underset{\varphi,\theta,\eta}{\text{min}} \left[ 1-\int \sqrt{P}\cdot\sqrt{\tilde{P}_\mathrm{rot}(\varphi,\theta,\eta)}\ dp_xdp_ydp_z \right]\ .
\end{align}
An optimization algorithm is used to find the chirality value $ \mu_\textrm{CV}$. For an achiral distribution $\mu_\textrm{CV}$ would be $0$. For a very chiral distribution the chirality value would be significantly different from $0$ (and might even be close to $1$). For the chirality value the three corresponding rotation angles are such that the overlap of $P$ and $\tilde{P}_\mathrm{rot}(\varphi,\theta,\eta)$ maximizes. The chirality value for the electron momentum distribution in Fig. \ref{fig4}(b) is $\mu_\textrm{CV}=0.09$. Fig. \ref{fig4}(c) shows $P-\tilde{P}_\mathrm{rot}'$ to visualize that the distribution in Fig. \ref{fig4}(b) is a chiral distribution. The red and blue isosurfaces correspond to the regions of $P$ and $\tilde{P}_\mathrm{rot}'$ that do not overlap. If $P$ were an achiral distribution the histogram in \ref{fig4}(c) would be empty.

The 3D histogram in Fig. \ref{fig4}(d) visualizes the presence of different laser electric fields and negative vector potentials in the entire focal volume. Since $\vec{E}(t)$ varies as a function of the position in the focal volume, a range of 3D fields contributes to the ionization probability when integrating over the focal volume. Fig. \ref{fig4}(d) visualizes the weighted occurrence of all laser electric fields within the focal volume. To this end, all electric field vectors are filled into a 3D histogram and weighted with the values from Fig. \ref{fig3}(b). The same is done for $-\vec{A}(t,\varphi_{ac})$. This procedure results in Fig. \ref{fig4}(d) showing which electric fields and which negative vector potentials mainly contribute to the overall ionization probability. The 3D isosurfaces show $\vec{E}(t)$ in orange and $-\vec{A}(t)$ in black, with the opacity of the isosurface indicating the relevance of the corresponding $\vec{E}(t)$ and $-\vec{A}(t)$.

Fig. \ref{fig4}(e) shows the total final electron momentum distribution, which is the sum of all electron momentum distributions for the ten different values of $\varphi_{ac}$ from Fig. \ref{fig3}(b), weighted with their approximated ionization probability $W_{\mathrm{SFA}}$. The distribution shown in Fig. \ref{fig4}(f) is generated the same way as Fig. \ref{fig4}(c) but quantifies the chirality of Fig. \ref{fig4}(e). The sum of all electron momentum distributions (see Fig. \ref{fig4}(e)) has a chirality value of $\mu_\textrm{CV}=0.11$. It is evident that the electron momentum distribution from the volume-averaged signal is still a chiral distribution.

The final momentum distribution shown in Fig. \ref{fig4}(e) is dominated by contributions from electric fields which have values of $\varphi_{ac}$ that are close to $2.6$ (see Fig. \ref{fig3}(b)). The field parameters in this work were chosen such that the Lissajous curve for the electric field with the highest ionization probability and the total final momentum distribution are both chiral. For a different set of field parameters, $W(\varphi_{ac})$ is not necessarily peaked at a single value of $\varphi_{ac}$, leading to a mixture of different 3D laser fields with similar ionization probability. However, there are many other sets of laser parameters that can be used to generate a well-defined chiral 3D field dominated by a single peak in the distribution $W(\varphi_{ac})$. Besides, it is also possible to generate well-defined 3D fields that produce electron momentum distributions which have very low chirality values.

\section{V. Conclusion}\label{sec:conclusion}

In conclusion, we present a simulation approach for strong-field ionization of single atoms in 3D laser fields and use it to demonstrate how to synthesize well-defined 3D laser fields that lead to chiral electron momentum distributions under realistic experimental conditions. Both the non-adiabatic tunneling dynamics and the Coulombic potential after tunneling are included in our simulation. The 3D laser fields are generated by overlapping two perpendicularly propagating two-color laser beams. The relative phase $\varphi_{ac}$ of the two beams is position-dependent and thus gives rise to a position-dependent 3D laser field. The variation of $\varphi_{ac}$ across the focal volume is taken into account for the calculation of the total final electron momentum distribution. Not all values of $\varphi_{ac}$ contribute equally, as the ionization probability varies significantly with $\varphi_{ac}$. Both the electron momentum distribution for the single 3D field with the highest ionization probability and the focal-volume-integrated distribution are chiral distributions for the field parameters considered in this work.

The proposed scheme establishes a benchmark for generating well-defined 3D laser fields under realistic experimental conditions and opens a path toward systematic studies of strong-field ionization in 3D light fields as a new class of experiments. Beyond fundamental interest, such 3D fields could enable all-optical enantiopurification \cite{neufeld2021strong}, the generation of chiral bound electronic states \cite{Mayer2022}, advances in laser-induced electron diffraction \cite{Huismans2011,Rajak2024}, and the investigation of sub-cycle interferences \cite{Meckel2008,Meckel2014,Xie2017,Eckart2018SubCycle}. Eventually, 3D laser fields may also be combined with pump-probe schemes using non-collinear laser beams \cite{moitra2025light}. Furthermore, the results of this work suggest that the angular distributions of emitted high harmonics in 3D laser fields may be influenced by the spatial grating ($W(\varphi_{ac})$) formed by the 3D field as well as by the 3D properties of the electron wave packet at the tunnel exit \cite{Murray2010, begin2020chiral, liu2016application}. Finally, the concept introduced in our manuscript could be used to design 3D light fields tailored for probing molecular structure and chirality via HHG, offering new possibilities for ultrafast imaging, as well as to control chemical reactions and electron dynamics harnessing the three-dimensionality of the driving laser field \cite{neufeld2021strong, ayuso2019synthetic, ordonez2023all, sun2025direct}.

\section{Acknowledgments}
Funded by the European Union. Views and opinions expressed are however those of the authors only and do not necessarily reflect those of the European Union or the European Research Council Executive Agency. Neither the European Union nor the granting authority can be held responsible for them. This work is supported by ERC-Starting Grant ``3DTunneling'' (101076166).

\section{Data Availability}
The data that support the findings of this article are openly available at \href{https://doi.org/10.5281/zenodo.15195910}{zenodo} \footnote{https://doi.org/10.5281/zenodo.15195910}.

\end{document}